# A Mobile Computing Architecture for Numerical Simulation

Cyril Dumont
*Paris 12 University*
*LACL*
*Créteil, France*
*dumont_cyril@yahoo.fr*

Fabrice Mourlin
*Paris 12 University*
*LACL*
*Créteil, France*
*fabrice.mourlin@wanadoo.fr*

**Abstract**

*The domain of numerical simulation is a place where the parallelization of numerical code is common. The definition of a numerical context means the configuration of resources such as memory, processor load and communication graph, with an evolving feature: the resources availability. A feature is often missing: the adaptability. It is not predictable and the adaptable aspect is essential. Without calling into question these implementations of these codes, we create an adaptive use of these implementations. Because the execution has to be driven by the availability of main resources, the components of a numeric computation have to react when their context changes. This paper offers a new architecture, a mobile computing architecture, based on mobile agents and JavaSpace. At the end of this paper, we apply our architecture to several case studies and obtain our first results.*

## 1. Introduction

The domain of numerical application is a place where technical frameworks are frequently used. There are useful to solve many practical problems, such as equation solvers, data distribution, and load observer. But the numerical simulations highlight software engineering problems, such as code parallelization, termination control, memory management, etc. Then, the simulation programmer has to change and rewrite numerical code for each experiment even if input data or computing resources are slightly changed. It seems obvious that this programming approach is expensive and it introduces new bugs and new cost.

A feature is clearly missing: adaptability. An adaptive computation tries to fulfill a set of goals in a complex, dynamic environment by sensing and acting upon its environment. In numerical simulation, the current constraints do not allow the realization of such computation. In this paper we present a new architecture dedicated to numerical adaptative computing.

For numerical application, the increased computational power of parallel hardware offers a promising way of exceeding limitation of resources. However, the parallelization of numerical code is usually an expensive task, especially when this task is performed from an existing code. For instance, numerical techniques such as Finite Difference Time Domain (FDTD) computer programs are used to analyze the electromagnetic environment in a lot of contexts. Implementations of this technique already exist such as GORFF-VE [1]. The development of a new parallel adaptive version is unbelievable but an adaptative use of this implementation is achievable. This kind of use allows two main features; the first one is a better use of the computing resources of the existing material architecture (large network or grid of processors). The second one is to allow the replay of experiment from a specific point, where a watch point is placed or where an error occurs.

In this paper, we present an architecture where mobility is the key concept to adapt all the data of a numerical experiment. These are the input data, the computing task and the material resources. The next section explains our technical choices and also how the mobility is exploited. The following part describes our software architecture and its evolution during a computing experiment. Next, we discuss advantages and limits of our approach, draw up its benchmark are built and finally we depict future directions of our activity.

## 2. Introduction to a dynamic network architecture

Traditional network architecture is static by nature. Network designers know in advance which computer hardware and software will participate in a specific computing and the network infrastructure is built in accordance with this concept. For example, in the Client/Server network architecture, specific server resources are designated to be served to clients on request, while other software units are indigenous on the clients.

In dynamic network architecture we do not decide in advance the specific hardware and software that will participate in the solution. Because of the multiplicity of software and hardware available on the network that could participate in the solution, it is advantageous to defer the decision until the software or hardware is actually required. The idea is to have the solution itself seek and recover on the network the components and resources it requires. Should the selected components and resources degrade or fail during execution, the solution can replace them or/and continue to operate.

## 2.1. A Mobile Agent Framework

Some programs can be made to run faster by dividing them up into smaller pieces and running these pieces on multiple processors.

For instance, an electromagnetic simulation about an electric circuit device to be analyzed (cable pigtail portion) starts with an input data set. This one describes an electric circuit device having a metal cabinet, all data input is described in terms familiar to professionals in the field but the size of this description is so large that it is not possible to read the whole set and to compute the field components on a single computer. A concrete experiment can use a 14 Go data set for a 1.4 GHz case study. Also the data set is partitioned with respect to the symmetries of the data. It means that the cut strategy depends on the symmetries which are express in the data set [2]. These properties can come from the underlying mesh if it is structured or from the future task, if that one is isotropic.

Several implementations already exist but few of them hide technical features from their environment. Data exchange is also a strong constraint in agent community; data type has to be preserved from the sender to the receivers. These main constraints helped us select a mobile agent framework.

## 2.2. The Jini Framework

**2.2.1. Jini introduction.** Our framework is based on Java language and Jini technology. The Jini networking system is a distributed infrastructure built around the Java programming language and environment. Jini is the name for a distributed infrastructure computing environment that can offer "network plug and play". A device or software service like an agent can be connected to a network and announce its presence, and clients that wish to use such a service can then locate it and call it to perform tasks [3].

The basic communication model is based on the semantic model of the Java Remote Method Invocation system, in which objects in one Java virtual machine communicate with objects in another one by receiving a proxy object that implements the same interface as the remote object. This communication model is the core feature for moving agents. The proxy object deals with all communication details between the two processes. The proxy object can introduce new code into the process to which it is moved. This is possible because Java byte codes are portable, and it is safe because of the Java environment's built-in verification and security.

**2.2.2. Jini Services.** The Jini technology includes several services. Both of them interest us particularly: JavaSpaces and Transaction Service.

Building distributed applications with conventional network tools usually entails passing messages between processes or invoking methods on remote objects. In JavaSpaces applications, in contrast, processes don't communicate directly, but instead coordinate their activities by exchanging objects through a space, or shared memory. A process can write new objects into a space, take objects from a space, or read (make a local copy of) objects in a space. When taking or reading objects, processes use simple matching, based on the values of fields, to find the objects that matter to them. If a matching object isn't found immediately, then a process can wait until one arrives. To modify an object, a process must explicitly remove it, update it, and reinsert it into the space. Spaces are object stores with several important properties that contribute to making JavaSpaces a powerful, expressive tool. Spaces are shared, persistent, transactionally secure [4].

Outrigger is the name of the contributed JavaSpaces(TM) service from Sun Microsystems. Transactions are a necessary part of many distributed operations. A series of operations, either within a single service or spanning multiple services, can be wrapped in a transaction. Mahalo is the transaction manager supplied by Sun as part of the Jini distribution. To make use of transactions, first it is necessary to create a transaction manager that can create and maintain a transaction for each client. This concept allows the user to ask for a set of requests which are closely related. It insures that the whole transaction is satisfied or cancelled. To locate a manager, you use Jini's lookup and discovery. Like all Jini services, the lookup service returns a proxy object to a transaction manager. The Jini transaction interfaces supply a service protocol needed to coordinate a two-phase commit.

## 3. Mobile computing architecture

A working space has to be organized, if we want to achieve a specific mission quickly. A JavaSpace follows this idea. In the Replicated-Worker pattern on JavaSpaces [5], also known as the Master-Worker pattern, a master process creates a collection of tasks that need to be run. Workers take tasks from the collection and run them, then hand over the computed result to the master. A space is a natural channel for passing messages between master and workers, due to the decoupled programming style it encourages.

Typically, there are many workers, and they are identical; hence the term replicated. This pattern neatly provides load balancing, whereby each worker contributes whatever resources it can afford. The worker on a faster machine will execute more tasks than the worker on a slower or otherwise heavily loaded machine; and as long as the granularity of the tasks is sufficiently fine, no worker will hold up the computation.

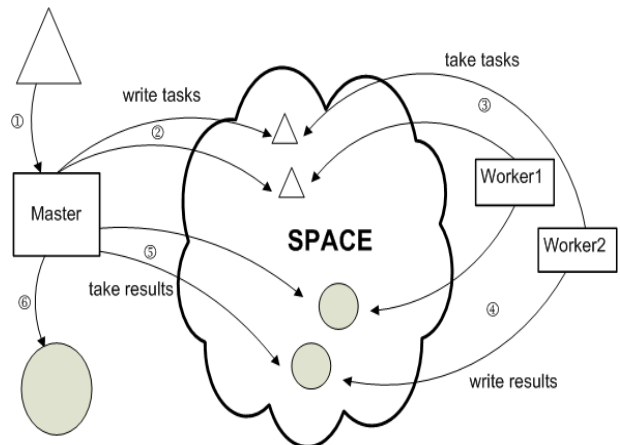

**Figure 1. The Replicated-Worker pattern on JavaSpaces**

The open source Java framework *ComputeFarm* [6] grew out of an implementation in JavaSpaces of this pattern. But our approach of this pattern differs from *ComputeFarm* because the tasks do not contain the required computation. The computation is on a mobile agent apart from the space. In the space, there are the parts of the file and the scheduler, which is the main component of the computation.

Based on the Replicated-Worker pattern, our mobile computing architecture goes further than the ComputeFarm framework. The adaptability and the replay of a computation case is the heart of our architecture.

In fact, our mobile computing architecture consists of four main components. All the components are on various nodes of a network. The link between all the components is the *Space*. Each computer containing a component must reach the computer containing the *Space*.

The *ComputingMaster* initializes the computation case and finishes it. It creates the parts of the source file and collects the files generated by the *ComputingWorker*.

The *ComputingAgent* is the mobile agent that contains the code of the computation case.

An essential part of the computation is the *Scheduler*, which schedules tasks for *ComputingWorker* to facilitate multitasking. When a component takes the *Scheduler*, the *Scheduler* is not available anymore. The component must rewrite it to make it available again. That will avoid conflicts and two executions of the same task. The strategy of choices of tasks takes part in the semantic of the *Scheduler*. The *Scheduler* we will use from now on attributes the tasks to the *ComputingWorker* in order of task arrival. Another implementation of the *Scheduler* is completely possible: priorities into the tasks for instance.

Finally the *ComputingWorker* take tasks from the *Scheduler* and with the *ComputingAgent* they can work with the parts of the file on the *Space*. Moreover, the result of a same task is identical whatever the ComputingWorker. The following part describes in detail the components in four steps.

### 3.1. Initialization step

Like the *TransactionManager* and the *JavaSpace*, the *ComputingAgent* is registered with the Jini lookup service: The lookup service, which is a Jini service, keeps track of the Jini services and provides proxies to communicate with the services.

This mobile agent is started up separately from the other components of the mobile computing architecture. It can be used by another computing case with another source file. Before starting the computation, it must be configured. A configuration file allows to define the following parts of the computation:
- The way to cut the source file and the number of parts.
- The number of *ComputingWorker* used for the beginning of the computation (this number can evolve with the passing of the computation).

- The address of the *ComputingAgent*. This address is used by the *ComputingWorker* to get the mobile agent.

With our configuration on hand, we can begin the computation. First of all, the set up of the *Scheduler* in the *Space* is essential. On the beginning it contains no task but it is the central point of the architecture. We said that the *ComputingMaster* ends the computation. For that it must listen to the *Scheduler*. When a task has been executed, the *ComputingMaster* wants to be informed. This component is the *ResultListener*.

Now we are ready to cut the source file. This part of the computation depends on the computation case (we have seen it in the configuration step). The several parts created are copied in a temporary directory. Execution of this part is completely asynchronous with the rest of the computation process. Fig. 2 describes the two steps of this initialization.

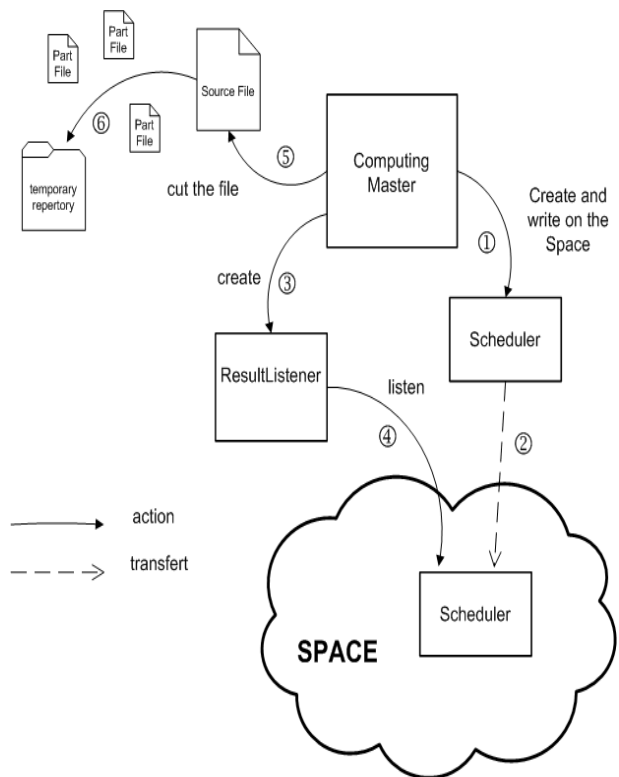

**Figure 2. Initialization step for a computation case**

The *ComputingMaster* creates other sub-components such as *FileEntryGenerator,* of which we will see the description in the following section.

### 3.2. Deployment step

In this section, we explain how a computing context can be initially deployed.

The first part of the deployment is to warn the *ComputingWorker* about the new computation. For that a mobile agent has its own road map, given by the ComputingMaster with the configuration file, and contains the address of the *ComputingAgent*. This mobile agent migrates on each computer of his roadmap and starts a *ComputingWorker* with the good configuration. Then, the ComputingMaster writes a *ConfigurationEntry* on the Space and each new ComputingWorker will be able to read the configuration.

We finished the previous section by the creation of the files and the fact that it is asynchronous. The *FileEntryGenerator* is created and can now wait for new file in the temporary directory. This allows, upon receipt of a file in this directory, to start the process of the generation of *FileEntry* and new *ComputingTask*.

A FileEntry is identified by an UUID (Universally Unique Identifier) used to identify the same entry in different contexts. A UUID is essentially a 16-byte (128-bit) number. In its canonical hexadecimal form a UUID may look like this: 550e8400-e29b-41d4-a716-446655440000. The number of theoretically possible UUIDs is therefore $2^{128} = 256^{16}$ or about $3.4 \times 10^{38}$. This means that 1 trillion UUIDs have to be created every nanosecond for 10 billion years to exhaust the number of UUIDs. With random UUIDs, the chance of two having the same value is poor.

It also contains data of part of the source file. Data are encoded Base64 to serialize data (mobile on the *Space)*. Before writing the entry on the *Space,* the *TranscationManager* creates a new transaction "A" that follows this part of the computing case during its various steps. This manager allows to check if the transaction is open.

With this transaction "A", the *FileEntry* is written on the Space and a *ComputingTask* is added on the *Scheduler*. The new task contains the transaction "A" and is in its first state: WAIT_FOR_COMPUTING. (Fig. 3)

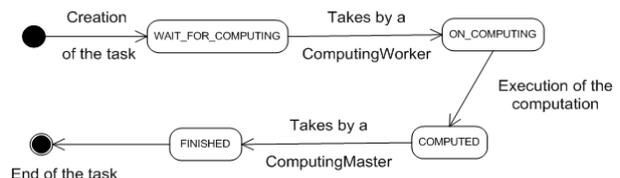

**Figure 3. Lifecycle of a ComputingTask**

### 3.3. Execution step

From now on the first *FileEntry* is written on the *Space,* the work of the *ComputingWorker* can begin.

Let us return to creation of the *ComputingWorker*, outlines quickly on the previous step. During its creation, the worker locates the *ComputingAgent* and calls it. Like all Jini services, the lookup service returns a proxy object to the agent. And this mobile agent (with its proxy) becomes the intelligence of the *ComputingWorker*.

To be informed on new *ComputingTask*, it must listen to the *Scheduler*. However a new task does not imply that it will be realized by this *ComputingWorker*. The lifecycle of a *ComputingWorker* is structured into two states (Fig. 4). The control of the activity of the *ComputingWorker* is essential; all resources of the *ComputingWorker* must be for the progress task and leave the others tasks for the ComputingWorker with unused resources. It can also accept task only if it is in WAIT_FOR_COMPUTING state.

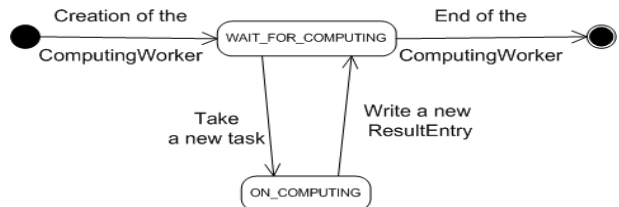

**Figure 4. Lifecycle of a ComputingWorker**

Now let us consider it is in this state and is informed that the *Scheduler* has a new *ComputingTask* in WAIT_FOR_COMPUTING state. The worker takes the *Scheduler* (① on Fig. 5), the new *ComputingTask* (② on Fig. 5) and analyses this task. Now it has a Transaction tnx1 to read a *FileEntry* on the *Space* (② on Fig. 5). With this entry, it decodes (Base64) data in a local directory (③ on Fig. 5) and starts the computation on the local file with the proxy of the *ComputingAgent* (④ on Fig. 5).

We put the *Scheduler* back on the *Space* so that other components of our architecture can listen to it. The state of the task is ON_COMPUTING (Fig. 3).

Then the file generated by the agent is transformed on a *ResultEntry* that has the same properties as *FileEntry*: a generated Uuid, data encoded in Base64 (⑤ on Fig. 5). This result is written on the *Space* with the transaction tnx1 (⑥ on Fig. 5). The ComputingTask is now in COMPUTED state (⑦ on Fig. 5).

This section explained the role of a unique *ComputingWorker*, but the principle of this architecture is the multiplicity of this kind of workers. The fundamental point is the possibility to add and/or remove any workers at any moment. This makes adaptability the main feature of our architecture. A new ComputingWorker can read the *Scheduler*, and if it does not have a configuration for this computation case, it can read the *ConfigurationEntry* on the *Space*. Now it can take a *ComputingTask* and executes it with the *ComputingAgent* (thanks to the new configuration).

### 3.4. Final step

A new *ResultEntry* was thus written on the *Space*. Now the *ResultListener*, which was created at the initialization step, will be useful. Like the *ComputingWorker*, the *ResultListener* must listen to the *Scheduler*. When a *ComputingTask* is in COMPUTED state (Fig. 3), the *ResultListener* is informed and can begin to create the result file. The *ResultListener* can begin to take the first results when the *ComputingMaster* is still cutting the source file.

A computation case is over when the number of the result files is equal to the number of the part of the source file. It is necessary to warn the *ComputingWorker* on this computation case. To do so, the *ComputingMaster* writes a new entry on the Space. This entry is a *StopEntry* and all *ComputingWorker* listen to the appearance of this kind of entry. When a *ComputingWorker* is informed, it stops its activity and is ready for a new computation case.

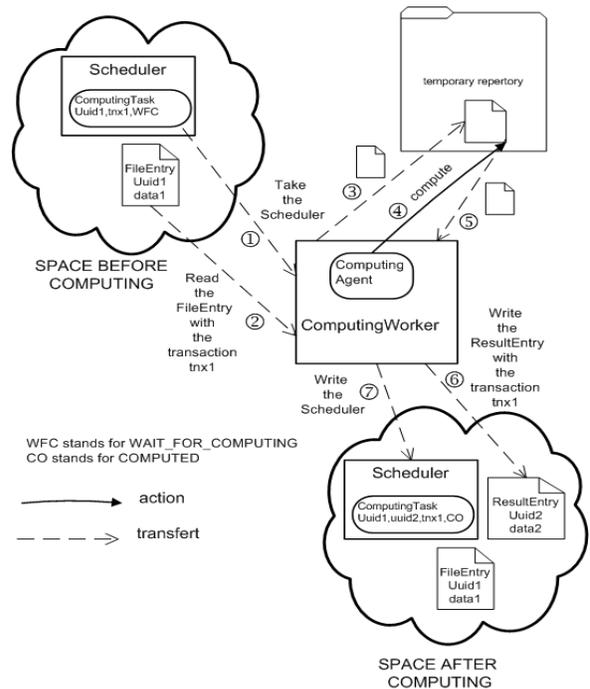

**Figure 5. The computing process**

The combination of two elements (transactions and the state of a *ComputingTask*) allows to start again a computation case. If a transaction is aborted, the task can start again from its last state and the *ComputingMaster*, with the *TransactionManager*, creates a new transaction for this task.

## 4. Synthesis on our approach

### 4.1. Observation on case study

We applied our architecture to several case studies. Of course, we used it for elementary numerical computation.

**4.1.1. The computation of the n'th decimal digit of various transcendental numbers.** For instance, we implement the formula for pi, discovered by David Bailey, Peter Borwein, and Simon Plouffe [7].

The reason this pi formula is so interesting is because it can be used to calculate the N-th digit of Pi (in base 16) without having to calculate all of the previous digits. Also, the computing task concept is clearly identified and the data set is light. A computing task is the computation of a particular component of the formula.

The cut file strategy is easy to implement. It depends on the number of decimal digits, which is useful to have by the end of the experiment.

Moreover, one can even do the calculation in a time that is essentially linear in N, with memory requirements only logarithmic in N. This is far better than previous algorithms for finding the N-th digit of Pi, which required keeping track of all the previous digits.

**4.1.2. A matrix computation.** A case study for computing the Cholesky decomposition of a positive definite matrix into a product of a lower and an upper triagonal matrix having a dimension of L (10 in our case) in a multi-channel procedure. In that case study, the data set is more important than before. All the matrixes are defined in an input file and the cut strategy isolates each subset of matrix [8].

The computation is performed in our parallel adaptive architecture comprising P processor modules (10 in our case); each processor module computes a new Cholesky decomposition from the entry file. This decomposition is divided into several computing tasks and because the computing tasks are not independent, the scheduling task which belongs to the Javaspace plays the role of "chief manager".

For each matrix, the scheduling consists in the computation of a set of rows of the lower triagonal matrix with R=L/P, whereby a given row j is assigned to a specific computing task of the Javaspace. The entries of the lower triagonal matrix are obtained by another computing task which contains an iteration process over the columns i, with i=1,...,L, whereby in each iteration step each processor module computes the entries of column i for its assigned rows R in a distinct computing task; with beginning an iteration step for computing the entries of column i with i≤L, the processor module to which the computation of row number i is assigned to stores its values of row i for access by the other processor modules.

This computing case brings more features about the context where our approach is more suitable. The definition of each kind of computing tasks is essential and especially the relation between all the computing tasks. This computing task development involves the cut file strategy. This allows more adaptive executions regarding the available processors.

### 4.2. Advantages and limits

In order to build our architecture, the Jini framework makes it easy to create dynamically networked components, applications, and services. Jini also allows an environment in which it is easy to create collaboration through the Jini Community. Jini moves the input data set as executable task entity of a JavaSpace executables. It means of a Java object over a network. The language is also known for its design flexibility. One of Jini's unique qualities is that it enables network self-healing and self-configuration, improving fault tolerance. Using Jini, networks easily adjust to changes in the computing environment.

Our experience highlights the importance of two main features: the definition of *ComputingAgent* and the decomposition of the input data. Both features involve adaptability of our system. In a more concrete context, *ComputingAgent* and input data already exist or are given by academic specialists. During an execution, the *ComputingWorker* receives *ComputingAgent* as a mobile agent which comes to realize its activity over an input data. Our architecture allows to provide a decomposition of the whole input data set as *FileEntry* (Fig. 5). This decomposition is a key feature to adapt the flow of activities to the available resource (*ComputingWorker* in Fig. 5).

### 4.3. Future direction

Our first results validate our approach and the adaptive architecture we presented earlier in this document. The four step experiment starts with initialization step which decides a lot of the constraints over all the execution. We work on possibilities of enlarging the choices such as the number of computing workers. Because new workers can be added into an experiment context, these new resources can be used.

Another direction is about comparison between our approach and existing parallel algorithms. We want to use Execution Time Measurement framework such as JETM [9]. Our objective is to create benchmark between MPI implementation of parallel algorithm and same implementation over our architecture.

## 5. Conclusion

This paper is about a new technique of resolution of linear systems in Java. Java, with Jini, can be use to develop distributed infrastructure with JavaSpace and transactions. That why we use this couple to make our architecture.

We stress the role of mobile agents for the *ComputingAgent* and also the role of the data decomposition which is closely related to the computing resource.

Our first results validate our design and they drive us to a more mobile approach where the decomposition will occur more than once. Also, this computing architecture allows to use existing numerical code as external *ComputingAgent*.

In our examples, only on *ComputingAgent,* with a numerical code, is used. In the case of a sequential computation, we must be able to make many computations one after the other. That why, the next evolution of our architecture is the complexity increase of our *Scheduler:* with a new definition of a *ComputingTask*. A ComputingTask will contain name of the *ComputingAgent* to call.